\documentclass{aastex62}

\received{April 24, 2019}
\accepted{November 6, 2019}
\shortauthors{Kang et al.}

\begin{document}

\title{Onset Mechanism of M6.5 Solar Flare Observed in Active Region 12371}

\correspondingauthor{Jihye Kang}
\email{siriustar@khu.ac.kr}

\author[0000-0002-0786-7307]{Jihye Kang}
\affil{Department of Astronomy and Space Science, Kyung Hee University, 1732 Deogyeong-daero, Giheung-gu, Yongin-si, Gyeonggi-do 17104, Korea}

\author{Satoshi Inoue}
\affiliation{Institute for Space-Earth Environmental Research, Nagoya University, Furo-cho, Chikusa-ku, Nagoya 464-8601, Japan}

\author{Kanya Kusano}
\affiliation{Institute for Space-Earth Environmental Research, Nagoya University, Furo-cho, Chikusa-ku, Nagoya 464-8601, Japan}

\author{Sung-Hong Park}
\affiliation{Institute for Space-Earth Environmental Research, Nagoya University, Furo-cho, Chikusa-ku, Nagoya 464-8601, Japan}

\author{Yong-Jae Moon}
\affiliation{Department of Astronomy and Space Science, Kyung Hee University, 1732 Deogyeong-daero, Giheung-gu, Yongin-si, Gyeonggi-do 17104, Korea}



\begin{abstract}

We studied a flare onset process in terms of stability of a three-dimensional (3D) magnetic field in active region 12371 producing an eruptive M6.5 flare in 2015 June 22. In order to reveal the 3D magnetic structure, we first  extrapolated the 3D coronal magnetic fields based on time series of the photospheric vector magnetic fields under a nonlinear force-free field (NLFFF) approximation. The NLFFFs nicely reproduced the observed sigmoidal structure which is widely considered to be preeruptive magnetic configuration. In particular, we found that the sigmoid is composed of two branches of sheared arcade loops. On the basis of the NLFFFs, we investigated the sheared arcade loops to explore the onset process of the eruptive flare using three representative magnetohydrodynamic instabilities: the kink, torus, and double arc instabilities (DAI). The DAI, recently proposed by Ishiguro $\&$ Kusano, is a double arc loop that can be more easily destabilized than a torus loop. Consequently, the NLFFFs are found to be quite stable against the kink and torus instabilities. However, the sheared arcade loops formed prior to the flare possibly become unstable against the DAI. As a possible scenario for the onset process of the M6.5 flare, we suggest a three-step process: (1) double arc loops are formed by the sheared arcade loops through the tether-cutting reconnection during an early phase of the flare, (2) the DAI contributes to the expansion of destabilized double arc loops, and (3) finally, the torus instability makes the full eruption.

\end{abstract}



\section{Introduction} \label{sec:intro}

The destabilization of the coronal magnetic field is important for the initiation of solar flares, filament eruptions, and also coronal mass ejections (CMEs). Many theoretical studies have been done in terms of the ideal magnetohydrodynamic (MHD) instability to reveal an onset process of the solar explosive phenomena. A magnetic flux rope (MFR), which is a cluster of twisted field lines, is often observed prior to the flares and it has been widely believed that the destabilization of the MFR drives the eruption.

One of the candidates for driving the eruptions is the kink instability \citep[KI;][]{1958PhFl....1..265K}. The KI can be excited if a winding number ($N$) of a field line around the magnetic axis of the MFR  exceeds a critical value, for instance, $N$=1.25 for a cylindrical MFR \citep{1979SoPh...64..303H} and $N$=1.75 for a semitorus type MFR \citep{2004A&A...413L..27T}. Several studies \citep[{\it e.g.},][]{2005ApJ...622L..69R, 2005ApJ...628L.163W, 2006ApJ...653..719A, 2012ApJ...756...59L, 2012ApJ...749...12Y, 2014A&A...572A..83K} reported that the KI drives the eruption of the MFR  and some of the MFRs could grow into CMEs. These highly twisted lines were supported by nonlinear force-free field (NLFFF) extrapolations in the several papers \citep{2014ApJ...780...55J, 2016ApJ...831L..18L, 2016ApJ...818..148L}. The torus instability \citep[TI;][]{1978mit..book.....B, 2006PhRvL..96y5002K}, which is also one of the ideal MHD instabilities, has been well investigated as an initiation of the MFR. The TI can occur when the force balance acting on the MFR is broken between the upward hoop force derived from the current flowing inside the MFR and downward strapping force derived from the external magnetic field. The TI is characterized by a decay index ($n$) value that measures how rapidly the external field above the MFR decreases. If the decay index reaches a critical value, for example, 1.5 in case of a semitorus MFR, it can be destabilized by the TI \citep{1978mit..book.....B, 2006PhRvL..96y5002K, 2010ApJ...718.1388D}. \citet{2018ApJ...864..138J} statistically examined the TI and the KI in both eruptive and confined events. They found that the TI plays an important role in distinguishing eruptive and confined events, but it is not a necessary condition for the events. They also found that the KI is not a major factor of eruptions.

Recently, another instability was newly proposed by \citet{2017ApJ...843..101I}, named as the double arc instability (DAI). They assumed that the sigmoidal structure often seen in the preeruptive stage is composed of the double arc loops and performed the stability analysis to the double arc magnetic configuration. They found that the double arc electric current loop can be destabilized even in a situation in which the TI is stable, for example, in the case in which the external field does not decrease (the decay index $n=0$). The critical condition of the DAI  is characterized by a $\kappa$ value, which is a magnetic twist and a ratio of the magnetic flux by the double arc loops, which are assumed to be formed through a reconnection of twisted field lines, and their overlying field lines.

In this paper, we first performed an extrapolation of the three-dimensional (3D) magnetic field  based on the photospheric vector magnetic field observed in solar active region (AR) 12371 via the NLFFF approximation. Next we conducted the stability analysis to the ideal MHD instabilities. Eventually we discuss the possible scenario for the flare triggering process. The organization of this paper is as follows. In section \ref{sec:method}, we present observations in AR 12371 and the numerical method for the NLFFF extrapolation. In section \ref{sec:results}, we show the NLFFF before the flare and results of the stabilities in a magnetic structure producing an M6.5 flare against the ideal MHD instabilities, such as the kink, torus and double arc. Then we discuss a possible scenario of the M6.5 flare and its eruption and conclude this paper in section \ref{sec:conclusion}.

\section{Observations and Numerical Method} \label{sec:method}

\begin{figure}[htbp]
\begin{center}
\plotone{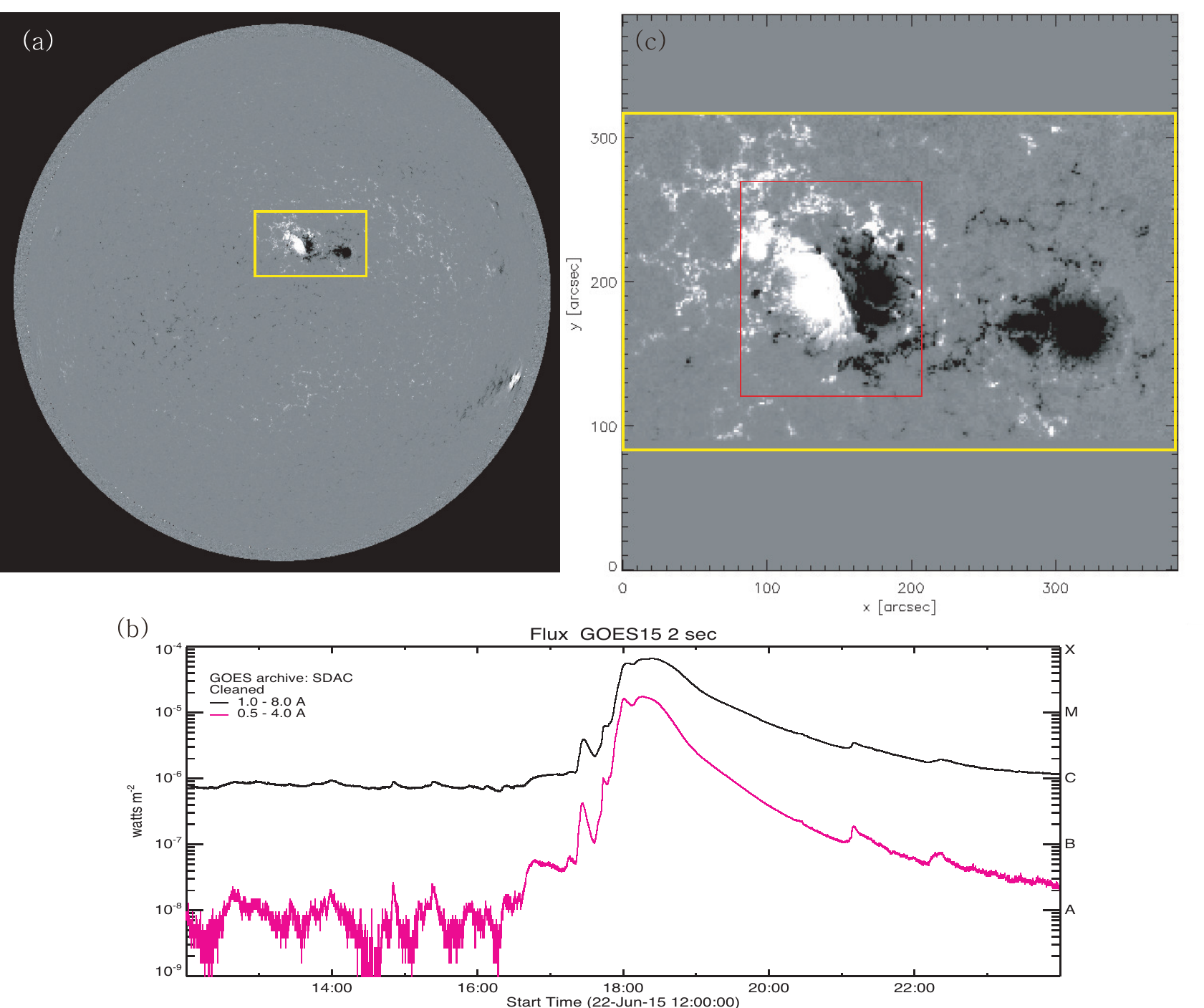}
\caption{(a) Full-disk magnetogram at 16:36 UT on 2015 June 22 observed by $SDO$/HMI. The SHARP data are available in a region enclosed by the yellow box. (b) Time profile of the X-ray flux measured by the GOES 12 satellite from 12:00 UT on 2015 June 22. The solar X-ray flux outputs in the 1$\sim$8\text{\AA} and 0.5$\sim$4\text{\AA} passbands are plotted, respectively. The M6.5 class flare is observed at 17:41 UT on 2015 June 22. (c) Bz distribution in AR 12371 at 16:36 UT used as the boundary condition of the NLFFF calculation. The SHARP data are set inside the yellow line and $\bold B \dbond \bold 0$ is assumed on the outside.
\label{fig:fig1}}
\end{center}
\end{figure}

 The AR 12371 is one of the very eruptive ARs and it produces major M-class flares accompanied with CMEs. During its disk passage, \citet{2017ApJ...845...59V} discovered that sigmoidal structures are formed before the major flares and shearless arcades are reformed after the major flares repeatably by tether-cutting reconnection. In this study, we focus on an eruptive event, the M6.5 flare, on 2015 June 22 when the AR 12371 was located near disk center. The AR consists of leading negative polarity spots and following spots of mixed polarities in which the M6.5 flare occurs. The M6.5 flare started at 17:39 UT and peaked at 18:23 UT in GOES soft X-ray flux at the eastern part of the AR 12371 (see Figures \ref{fig:fig1}(a) and (b)). This flare is also associated with a halo CME. \citet{2017ApJ...845...59V} showed large shearing and converging motions in the following polarity sunspot before the flare. Rotational and shear flows are also found on both sides of the PIL during the flare \citep{2017ApJ...849L..35B, 2018ApJ...853..143W}. The Helioseismic and Magnetic Imager \citep[HMI;][]{2012SoPh..275..207S} on board the Solar Dynamics Observatory \citep[\it SDO;][]{2012SoPh..275....3P} provides a time series of vector magnetic fields that are formatted in the Spaceweather HMI Active Region Patch \citep[SHARP;][]{2014SoPh..289.3549B}. Figure \ref{fig:fig1}(c) shows the $B_z$ distribution taken at 16:36 UT. The NLFFF calculations are conducted using the SHARP data observed at 13:36, 14:36, 15:36, and 16:36 UT before the flare and 18:36 and 19:36 UT after the flare. The region of the original SHARP data is covered by 474 $\arcsec$$\times$ 226.$\arcsec$5 resolved by 948 $\times$ 453 $px^2$. In this study, however, we reduced the resolution through a 3 $\times$ 3 binning process with  a resolution of 316 $\times$ 151 as seen in Figure 1(c). For the NLFFF calculations, the SHARP data reduce in the horizontal direction and extend in the vertical direction to 256 $\times$ 256 uniform grid points, which are totally assigned in Cartesian coordinates, because the original SHARP data do not fully cover the observed coronal structure structure in the AR 12371. The extended regions are assumed to have no magnetic field, i.e., $\bold B \dbond \bold 0$.

The NLFFF extrapolation was performed based on the MHD relaxation method developed by \citet{2014ApJ...780..101I}. This method seeks the force-free solution satisfying the given photospheric vector magnetic field by solving the zero-beta MHD equations as follows (dimensionless form) :
\begin{equation}
\rho \dbond \left\vert \bf B \right\vert,
\end{equation}
\begin{equation}
{\partial  \bf v \over \partial \mit t}
\dbond -(\bf v \bf \cdot \bf \nabla)\bf v
+{1 \over \rho}  \bf  \bf j \times \bf B
+\nu \nabla^2 \bf v,
\end{equation}
\begin{equation}
{\partial \bf B \over \partial \mit t}
 \dbond \nabla \times \left ( \bf v \times \bf B -\eta j \right ) - \nabla \phi,
\end{equation}
\begin{equation}
\bf j \dbond \nabla \times \bf B,
\end{equation}
and
\begin{equation}
{\partial \phi \over \partial \mit t} + c_h^2 \nabla \cdot \bf B \dbond -{\mit c_h^2 \over c_p^2} \phi,
\label{eq:dedner}
\end{equation}
where $\rho$, $\bf B$, $\bf v$, $\bf j$ and $\phi$ represent the hypothetical gas density, magnetic field, plasma flow velocity, electric current density, and a scalar potential function, respectively. Equation (\ref{eq:dedner}) is used to reduce derivation from $\nabla \cdot {\bf B} =0$ for the divergence-free condition \citep{2002JCoPh.175..645D}. We used the same constant parameters as those set in \citet{2016PASJ...68..101K} except the resistivity $\eta$. The resistivity follows as
\begin{equation}
\eta = \eta_0
+ \eta_1 {\left\vert \bf j \times \bf B\right\vert \left\vert \bf v \right\vert^2 \over \left\vert \bf B \right\vert ^2},
\label{eq:eta}
\end{equation}
where $\eta_0 = 10^{-5}$ and $\eta_1 = 7\times10^{-3}$. The second term plays a role in quickly accelerating the magnetic field toward the force-free state. The SHARP data are used for the bottom boundary conditions of the NLFFF calculations. The potential field derived from the Green's function method \citep{1982SoPh...76..301S} is given as an initial state for the NLFFF calculation. Regarding to the boundary condition, the potential field is fixed at the lateral and top boundaries during the calculation while the observed magnetic field is set at the bottom boundary. In order to seek force-free fields by this method, we gradually change the horizontal components of magnetic fields on the bottom from the potential fields to observational fields, according to the MHD process but neglecting the pressure and gravity. After the boundary data completely fits to the observational ones, the iteration is further performed until the solution converges to an equilibrium state. Since the solution is obtained through the zero-$\beta$ MHD process, the equilibrium solution corresponds to the force-free fields. A detailed process of this method was described in \citet{2014ApJ...780..101I} and \citet{2016PEPS....3...19I}.

\section{Results} \label{sec:results}

\subsection{NLFFF Before and After Onset of the M6.5 Flare} \label{subsec:magnetic}

\begin{figure}[htbp]
\begin{center}
\plotone{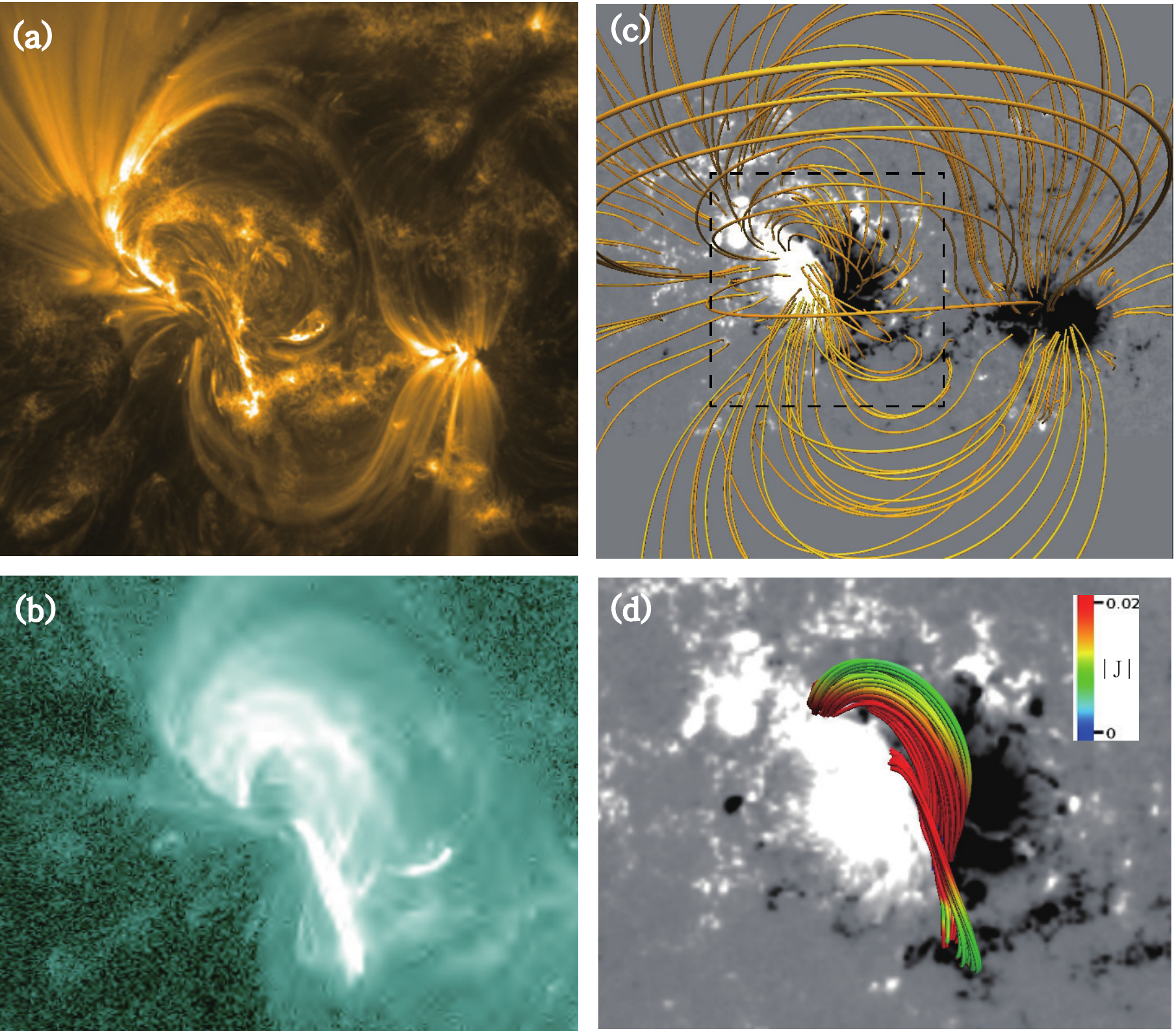}
\caption{(a), (b) EUV 171\text{\AA} and 94\text{\AA} images at 16:36 UT (approximately an hour before the M6.5 flare) taken by $\it SDO$/AIA, respectively. (c) Overall magnetic field lines obtained from the NLFFF model. (d) Magnetic field lines close to the main polarity inversion line. Field line color indicates the strength of the current density.
\label{fig:fig2}}
\end{center}
\end{figure}

In order to compare with the NLFFFs, we used extreme ultraviolet (EUV) images observed on 2015 June 22 16:36 UT taken approximately an hour prior to the flare onset time from  Atmospheric Imaging Assembly \citep[AIA;][]{2012SoPh..275...17L} on board {\it SDO}. Figures \ref{fig:fig2} (a) and (b) indicate the AIA 171\text{\AA} and 94 \text{\AA} images which show overall the magnetic and sigmoidal structures clearly. Figure \ref{fig:fig2}(c) shows magnetic field lines obtained from the NLFFF at the same time. Those field lines nicely capture the observed overall coronal loops shown in the SDO/AIA 171\text{\AA} image. Figure \ref{fig:fig2}(b) shows the high temperature plasma distribution which corresponds to the highly sheared field lines where the strong current density is enhanced. The NLFFFs shown in Figure \ref{fig:fig2}(d) also capture these highly sheared field lines well. The highly sheared field lines are also reconstructed by the optimization procedure and are consistent with an observed inverse-S shaped structure \citep{2018ApJ...857...90V}.

\subsection{Stability of the NLFFF} \label{subsec:stability}

We next diagnose the stability of the NLFFFs against the ideal MHD instabilities because those are possible candidates driving the solar eruptions \citep{2016PEPS....3...19I, 2018SSRv..214...46G}. We first calculate a magnetic twist to investigate the stability of the KI. We estimated the magnetic twist according to \citet{2006JPhA...39.8321B}
\begin{equation}
T_w \dbond {1 \over 4\pi}\int \frac{{\bf \nabla}\times \bf B \cdot  \bf B}{|\bf B|^2} dl,
\end{equation}
where $dl$ is a line element.

\begin{figure}[htbp]
\begin{center}
\plotone{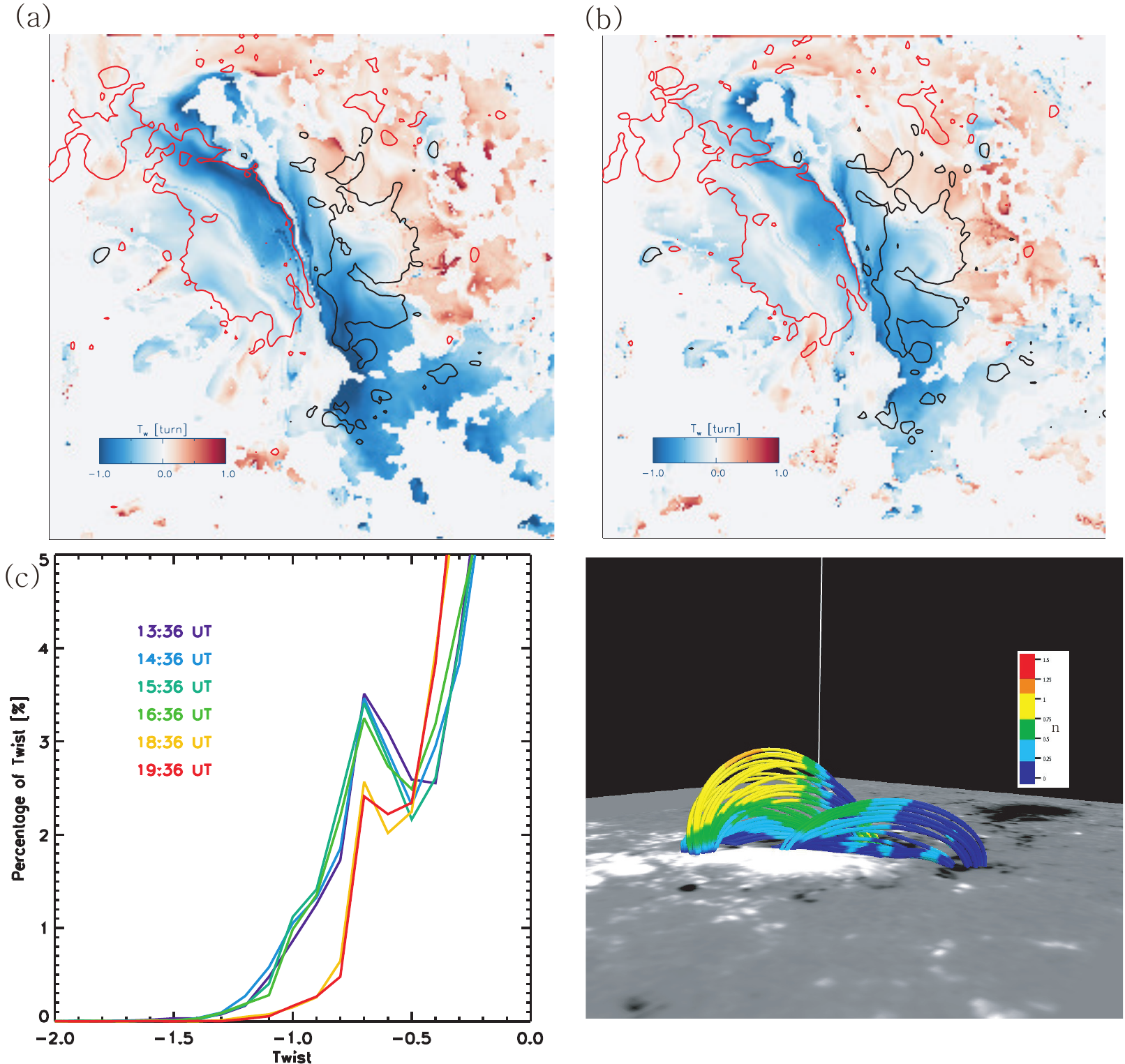}
\caption{(a), (b) Distribution of the magnetic twist obtained from the NLFFF at 16:36 UT (approximately an hour before the flare) and at 18:36 UT (approximately an hour after the flare), respectively. The magnetic twists are calculated within a dashed box shown in Figure \ref{fig:fig2}(c). Red and blue lines indicate contours of $B_z=500 (G)$ and $B_z=-500 (G)$, respectively. (c) Time series of histograms of the magnetic twist before (13:36, 14:36, 15:36, and 16:36 UT) and after (18:36 and 19:36 UT) the flare. The values are normalized by a total grid number of the region in the panel (a). (d) Side view of the magnetic field lines close to the main PIL at 16:36 UT. The selected field lines are the same as those in Figure \ref{fig:fig2}(d) and their descritized color indicates the value of the decay index.
\label{fig:fig3}}
\end{center}
\end{figure}

Figure \ref{fig:fig3}(a) shows the magnetic twist distribution with $B_z=500G$ (red line) and $B_z=-500G$ (black line) approximately an hour before the flare onset time (16:36 UT). The magnetic twist is measured by the region the $|B_z| > 25G$ within a dashed box shown in Figure \ref{fig:fig2}(c) which covers the main bipolar region producing the flare. The magnetic twist is dominated by a negative sign near the main PIL producing the flare, and surrounding regions are occupied by relatively weak positive twist before and after the flare. The distribution of magnetic twist obtained from this study is similar to that of \citet{2018ApJ...857...90V}. Although strongly twisted lines with more than one turn are found, moderately twisted lines ranging from 0.5 to 1.0 dominate over the region close to the PIL. On the other hand, these twisted field lines are significantly relaxed after the flare onset time (18:36 UT) as shown in Figure \ref{fig:fig3}(b). Figure \ref{fig:fig3}(c) shows histograms of the magnetic twist normalized by a total grid number of the region where we measured the magnetic twist (the dashed box in Figure \ref{fig:fig2}(c)). We found the same tendencies ,i.e., the  twisted lines, in particular, with more than a half-turn twist, formed before the flare obviously disappeared after the flare. As we focus on the moderately twisted lines, the magnetic fields are twisted near the flare onset time. This result suggests that the twisted lines with more than a half-turn play an important role in driving the M6.5 flare.  Although the magnetic twist is gradually accumulated toward the flare onset time, we could not find strongly twisted lines causing the KI \citep[$T_w=1.75;$][]{2004A&A...413L..27T}.

We also diagnose the stability to the TI. The decay index value is defined as
\begin{equation}
n \tbond - {{d \log |\bf B_{p,t}|} \over {d \log z}},
\end{equation}
where $\bold{B_{p,t}}$ is the horizontal component of the external field, which is assumed as a potential field. The decay index is how rapid the horizontal component of the magnetic field decreases with height and is well known as a parameter determining the TI. \citet{2006PhRvL..96y5002K} and \citet{2007AN....328..743T} pointed out that the TI is excited if $n\ge1.5$ is satisfied. The critical limit of the decay index has been debated because it strongly depends on the boundary condition and the shape of a flux rope \citep{2015Natur.528..526M}. \citet{2010ApJ...718..433O} pointed out a lower critical decay index, 1.42, by applying the flux rope separation. \citet{2015ApJ...814..126Z, 2016ApJ...821L..23Z} suggested that the critical decay index estimated in their own flux rope models gets lower than 1.5 and even down to 1.1. Figure \ref{fig:fig3}(d) shows the twisted lines colored in a value of the decay index. An orange color scale shown in Figure \ref{fig:fig3}(d) indicates that the decay index is larger than 1.1 which is the lower limit of \citet{2015ApJ...814..126Z, 2016ApJ...821L..23Z}. We found the small orange region whose decay index is larger than 1.1 in top of the twisted field lines. According to this result, most of the twisted lines do not reach the area where $n\ge 1.1$ satisfies, so that the magnetic field is suggested to be stable against the TI. Therefore, another mechanism is required to drive the eruption.

\subsection{Stability to the DAI} \label{subsec:DAI}

We investigate the DAI recently proposed by \citet{2017ApJ...843..101I}. They pointed out that the double arc loops, in particular, the joint region bridging two sheared loops becomes unstable. This instability can well explain a reason why the tether-cutting reconnection \citep{2001ApJ...552..833M} can drive an eruption. This instability is characterized by a $\kappa$ parameter which is defined as,
\begin{equation}
\kappa \tbond T_w {\Phi_{rec} \over \Phi_{tot}} \end{equation}
where $\Phi_{rec}$ is a magnetic flux of the double arc loops and $\Phi_{tot}$ is a total flux of the field lines surrounding the double arc loops, respectively. It is, however, hard to measure the kappa ($\kappa$) from observations, because of a lack of information on the magnetic twist of field lines and the reconnected flux. Therefore, we first use the $\kappa^*$ according to \citet{2018ApJ...863..162M},
\begin{equation}
\kappa^* \approx {\int T_w |B_r| \mathrm{d}S \over \Phi_{tot}}.
\label{eq:k*}
\end{equation}
In this formula, the reconnected flux is assumed to an integration of the magnetic twist as a function of magnetic flux ($B_z$). \citet{2018ApJ...863..162M} estimated the $\kappa^*_{T_c}$ in the AR 11158 for 2 days using equation (\ref{eq:k*}) assuming that the double arc loops (flux rope) are formed through a reconnection between the twisted lines with more than a critical twist ($T_c$). Therefore, the reconnected flux in their study was estimated using the flux occupied with twisted lines with more than the $T_c$. They also calculated the total flux which is the magnetic flux of the overlying field lines crossing above the main PIL. Their results show that a {$\kappa^*_{0.5}$} assuming that the reconnection takes place in the twisted lines with more than a half-turn slightly increases before the flare occurrence and decreases after the flare onset. The temporal evolution of a {$\kappa^*_{0.5}$ in the AR 12371 following the method of \citet{2018ApJ...863..162M} is similar to the one obtained in the AR 11158 while the $\kappa^*_{0.5}$ is larger than theirs. This is the reason why the NLFFFs in the AR 12371 have more field lines with twist larger than a half-turn compared to those of the AR 11158 in Figure \ref{fig:fig4}(a). However, there is no guarantee that the all of those twisted lines participate in the reconnection to drive the solar flare.

 In order to calculate the {$\kappa^*$} value carefully, we detect the field lines experiencing the reconnection in the early stage of the flare. According to the standard flare model \citep{2011LRSP....8....6S}, the flare ribbons correspond to a region where the footpoints of the reconnected field lines are anchored. Therefore, we can expect that the reconstructed field lines before and after the flare traced from the flare ribbons would participate or have participated in the reconnection. We use the 1600 \text{\AA} images of the upper photosphere obtained from {\it SDO}/AIA to clearly detect the two flare ribbons.  We integrate the flaring ribbon region during 10 minutes from an initial brightening at 17:41 UT to just prior to a separation of the flare ribbons at 17:51 UT. We only consider the initial brightening of the flare ribbons showing an elongation motion because the double arc loops would be formed in this phase. Figures \ref{fig:fig4} (a) and (b) show the magnetic twist distribution (gray scale map) before and after the flare and the location of the integrated flare ribbons (green line), respectively. The integration of brightening for the flare ribbons during the initial phase (inside of the green contour) covers the footpoints of the strongly twisted field lines, in Figure \ref{fig:fig4}(b), while these twists are relaxed after the flare. Figure \ref{fig:fig4}(c) shows the results of the field lines traced from the flare ribbons. This configuration looks to be double arc field lines where the strong current density is enhanced. Therefore, the DAI might be a possible candidate for the eruption. We determined the total flux, which is the overlying magnetic flux only passing over the main PIL within the dashed box shown in Figure \ref{fig:fig2}(c).

Figure \ref{fig:fig4}(d) shows results of the $\kappa^*$ (black line with cross sign) measured from our approach and the $\kappa^*_{0.5}$. (blue line with triangle sign) following \citet{2018ApJ...863..162M} in which the double arc loops are selected by the twisted field lines with more than a half-turn before and after the flare onset. We only consider the $\kappa^*$ approximately an hour before and after the flare onset time because the magnetic structure close to the flare occurrence time is directly related to the flare ribbons. Both of them exceeds 0.1, which is a threshold of the DAI, while the $\kappa^*$ is almost half in comparison with the $\kappa^*_{0.5}$. Even in a more strict situation, we found that the $\kappa^*$ exceeds 0.1. After the flare, both of them ($\kappa^*$ and $\kappa^*_{0.5}$) fall down to the value below the threshold.

\begin{figure}[htbp]
\begin{center}
\plotone{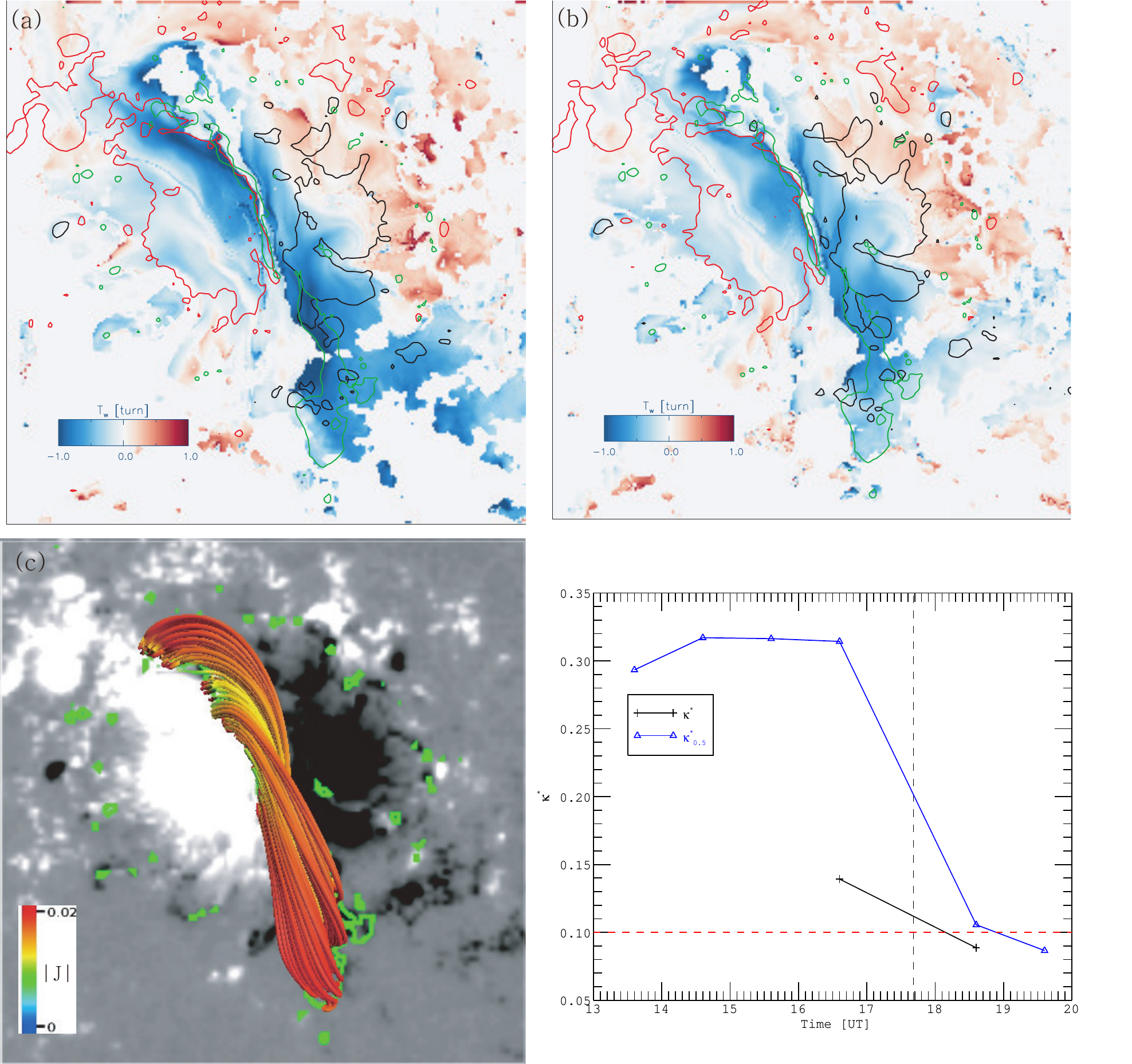}
\caption{(a), (b) The same format in figures \ref{fig:fig3}(a) and (b) except that the integrated flare ribbon (green contour appeared) during the initial phase of the flare is plotted from 17:41 UT to 17:51 UT. (c) The twisted field lines rooted on the integrated ribbons (inside of green contour). The color of field lines indicates the electric current density. (d) Temporal evolution of the $\kappa^*$ (black line with cross symbol) estimated from the twisted lines as shown in (c) and $\kappa^*_{0.5}$ (blue line with triangle sign) derived from the method of \citet{2018ApJ...863..162M} assuming that the reconnection took place between the twisted field lines with more than the half-turn. Vertical black dashed line and red dashed line indicate the flare onset time and the critical threshold of the kappa ($\kappa$) derived from \citet{2017ApJ...843..101I}.
\label{fig:fig4}}
\end{center}
\end{figure}

\section{Discussion and Conclusion} \label{sec:conclusion}

We have examined the stability of the magnetic field against the ideal MHD instability, such as the kink, torus, and double arc instabilities, before the M6.5 flare on 2015 June 22 in AR 12371 using the time series of the NLFFFs based on the photospheric vector magnetic fields.

According to our analysis, most of the twisted lines found in the NLFFFs are less than one turn before the M6.5 flare and significantly relaxed after the flare. We first suggested that the magnetic fields are stable to the KI because the twisted field lines do not reach the threshold value causing the instability. Furthermore, we also found that the top of the twisted lines is below a height where the TI is excited. Since the halo CME was observed to be associated with the M6.5 flare, some processes are required to lift the twisted lines up to the area where those become unstable to the TI.  We found that the two branches of sheared arcade loops are possibly unstable against the DAI while it is stable against the TI in the AR 12371.

 The sheared arcade loops are formed through the helicity injection derived from the long-term evolution for a few days prior to the flare \citep{2017ApJ...845...59V}. Because the sheared arcade loops are quite stable, some triggering processes are required to make them unstable. \citealt{2017NatAs...1E..85W} found small islands having opposite polarities to the net flux on the main PIL before the flare using the Near-InfraRed Imaging Spectropolarimeter \citep{2012ASPC..463..291C} of the 1.6m Goode Solar Telescope \citep[GST;][]{2010AN....331..620G}. An island, which is a counterpart of the island observed by GST, is located around an intersection region between the two branches of sheared arcade loops shown in Figure \ref{fig:fig5} (a). The island appeared in the {\it SDO}/HMI around 15:24 UT which is approximately two and half hours before the flare occurrence time and its size is less than 3$\arcsec$ shown in Figures \ref{fig:fig5} (a) and (b). It, furthermore, becomes bigger toward the flare onset time. In addition, we found that the strong vertical electric current density ($j_z$) appears inside the island at the same time and it is enhanced toward the flare onset time as  seen in Figure \ref{fig:fig5} (b). This would be consistent with a flare triggering model proposed by \citet{2012ApJ...760...31K}. According to their model, the reconnection between the two branches of sheared arcade loops  and the small island can make a large flux rope leading to the eruption. Therefore, the small island might play a key role in accelerating the magnetic reconnection, especially the tether-cutting reconnection. \citet{2017ApJ...845...59V} also suggested that the erupting flux rope should be formed by the tether-cutting reconnection. Figure \ref{fig:fig6} shows the temporal evolution of the small island. The magnetic flux increases toward the flare onset time. On the other hand, the small island  could be captured with only a few grids, which is a strong limitation of spatial resolution of the {\it SDO}/HMI. Therefore, we need more high-resolution observations for detailed analysis.

 As one of possible scenarios of the M6.5 flare and the eruption, we suggest the following three-step process. First, the tether-cutting reconnection between the two branches of sheared arcade loops leads to a formation of the double arc loops and accumulating magnetic twist during the early stage of the flare. The tether-cutting reconnection plays a role in transforming the sheared arcade loops into the the double arc loops whose configuration is a key criterion to estimate the DAI. In the second stage, once the double arc loops become unstable against the DAI, the rising and expanding loops change into the torus shaped loops even though they are still stable against the TI.  At a final stage, when the torus loops reach to the region where TI can take place, they can finally undergo the full eruption. Therefore, multiple processes including the tether-cutting reconnection, DAI, and TI cause the eruptive flare. In particular, the DAI plays an important role in raising the transformed torus loops to the height where the TI takes place \citep{2017ApJ...843..101I, 2018ApJ...857...90V}.

The twist and the decay index of magnetic field lines have been studied as a key factor for the KI and the TI, respectively. For example, calculating normalized helicity flux in AR 12371, \citet{2017ApJ...845...59V} found that it shows gradual increases over a few days before the flare. A major difference of the parameter $\kappa$ studied here is that it takes into account not only the magnetic twist but also the reconnection and overlying fluxes. This may help to reveal the onset mechanism of the M6.5 flare, which is presumed to be triggered by the DAI.

In order to contribute to the spaceweather forecast using this method, we need to detect the flare ribbons before the occurrence of the solar flare. It is, unfortunately, very difficult to know even the basic information of the flare ribbons beforehand. We require further improvements to calculate the $\kappa^*$ in the preflare phase.

\begin{figure}[htbp]
\begin{center}
\plotone{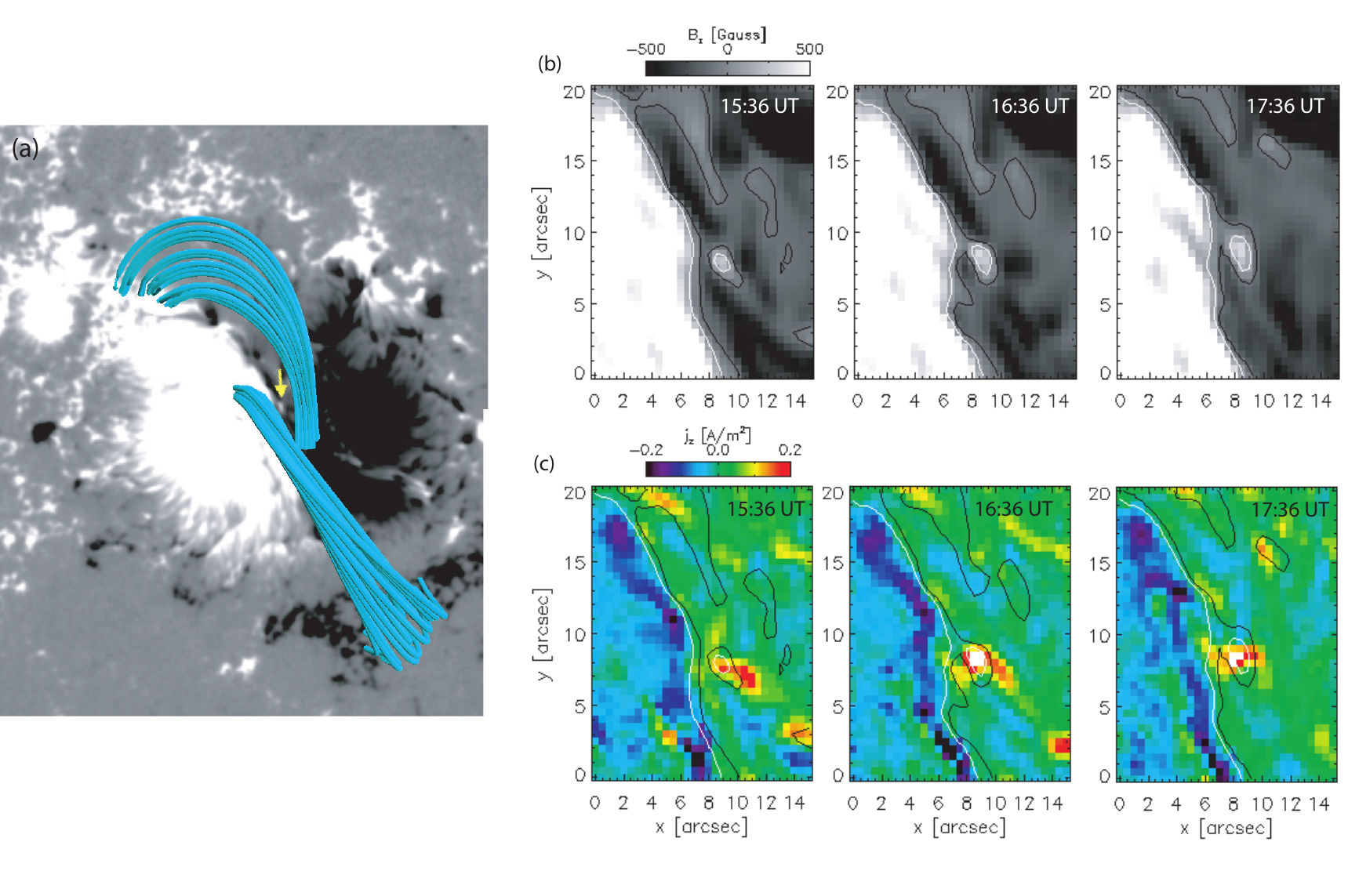}
\caption{(a) $B_z$ distribution at 16:36 UT in the bipolar region surrounded by the square in Figure \ref{fig:fig1}. Yellow arrow indicates the small island near the PIL and its neighboring field lines are drawn in cyan. (b), (c) Temporal evolution of $B_z$ and $j_z$ distributions in an enlarged region including the small island, respectively. White and black lines indicate $B_z=100(G)$ and $B_z=-100(G)$, respectively.
\label{fig:fig5}}
\end{center}
\end{figure}

\begin{figure}[htbp]
\begin{center}
\plotone{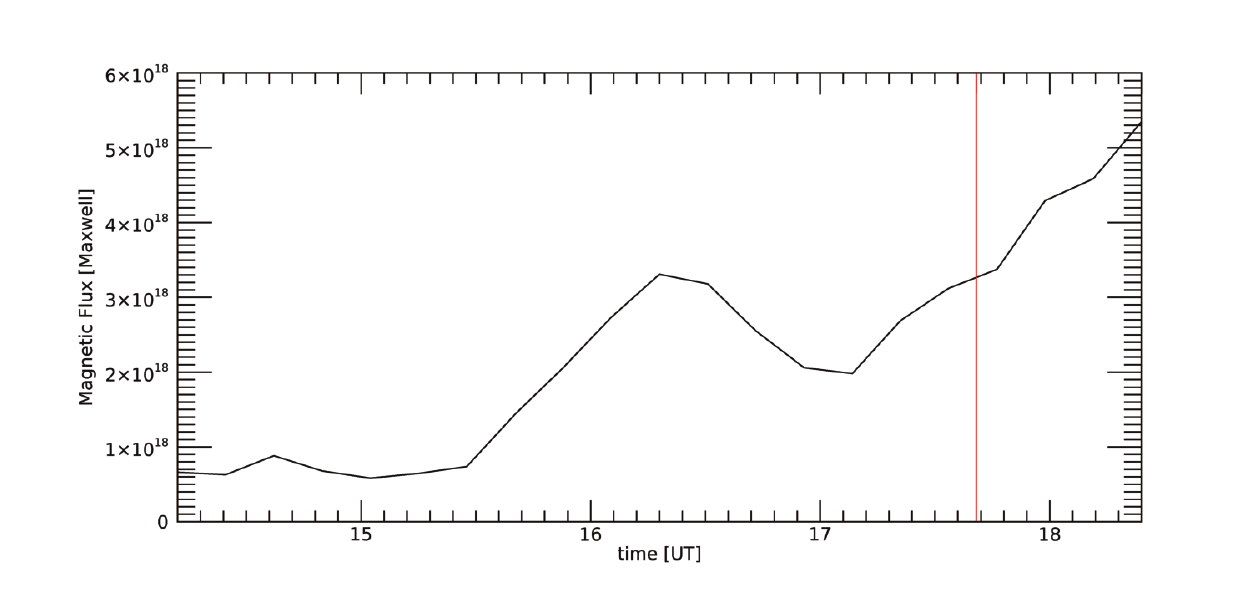}
\caption{Temporal evolution of the magnetic flux of the small island. The red line indicates the flare onset time.
\label{fig:fig6}}
\end{center}
\end{figure}

\acknowledgments
 The authors are grateful to the anonymous referee for constructive comments that improved the paper. J.K. thanks to the Kyung Hee University for general support of this research. This research was supported by Basic Science Research Program through the National Research Foundation of Korea (NRF) funded by the Ministry to Education (2017R1A6A3A11027930), as well as by BK21 plus program through the NRF. This study was partially supported by an Institute for Information $\&$ communications Technology Promotion (IITP) grant funded by the Korea government (MSIP) (2018-0-01422, Study on analysis and prediction technique of solar flares) and JSPS KAKENHI grant Nos. JP15H05814 and MEXT as "Exploratory Challenge on Post-K Computer"(Environmental Variations of Planets in the Solar System). Figures \ref{fig:fig2} (b), (d) and Figure \ref{fig:fig3} (d) were visualized by VAPOR, which is a product of the National Center for Atmospheric Research's Computational and Information Systems Lab \citep{2005SPIE.5669..284C, 2007NJPh....9..301C}.

\end{document}